\documentclass[journal]{IEEEtran}
\usepackage{graphicx}
\usepackage{algorithm}
\usepackage[noend]{algpseudocode}
\usepackage{caption}
\usepackage{cite}
\usepackage{url}
\begin{document}
\title{An energy efficient routing protocol for wireless Internet-of-Things sensor networks}

\author{\IEEEauthorblockN{Vidushi~Vashishth\IEEEauthorrefmark{3},
        Anshuman~Chhabra\IEEEauthorrefmark{2}\IEEEauthorrefmark{4},
        Anirudh~Khanna\IEEEauthorrefmark{2},
        Deepak~Kumar~Sharma\textsuperscript{*}\IEEEauthorrefmark{3}\thanks{* Corresponding Author},
        Jyotsna~Singh\IEEEauthorrefmark{2}\\[1.5ex]}
        
\IEEEauthorblockA{\IEEEauthorrefmark{2}Division of Electronics and Communication Engineering\\
Netaji Subhas Institute of Technology, University of Delhi\\
New Delhi, India\\
\emph{Email: anirudhkhanna9654@gmail.com, jsingh.nsit@gmail.com}\\[1.5ex]
\IEEEauthorrefmark{3}Division of Information Technology\\
Netaji Subhas Institute of Technology, University of Delhi\\
New Delhi, India\\
\emph{Email: vidushivashishth96@gmail.com, dk.sharma1982@yahoo.com}\\[1.5ex]
\IEEEauthorrefmark{4}Department of Computer Science\\
University of California, Davis\\
New Delhi, India\\
\emph{Email: chhabra@ucdavis.edu}\\
}}

\iffalse
\author{\textbf{BTP Project 2018}:
Anshuman~Chhabra,
        Vidushi~Vashishth,
        Anirudh~Khanna\\
        \textbf{Supervisors}:  Dr. Jyotsna~Singh and
        Dr. Deepak~Kumar~Sharma% <-this % stops a space
}
\markboth{IEEE Internet of Things Journal}%
{Shell \MakeLowercase{\textit{et al.}}: Bare Demo of IEEEtran.cls for IEEE Journals}
\fi


\maketitle

\begin{abstract}
Internet of Things (IoT) are increasingly being adopted into practical applications such as security systems, smart infrastructure, traffic management, weather systems, among others. While the scale of these applications is enormous, device capabilities, particularly in terms of battery life and energy efficiency are limited. Despite research being done to ameliorate these shortcomings, wireless IoT networks still cannot guarantee satisfactory network lifetimes and prolonged sensing coverage. Moreover, proposed schemes in literature are convoluted and cannot be easily implemented in real-world scenarios. This necessitates the development of a simple yet energy efficient routing scheme for wireless IoT sensor networks. This paper models the energy constraint problem of devices in IoT applications as an optimization problem. To conserve energy of devices the proposed protocol makes use of clustering, cluster head election and least energy-expensive path computation for efficient and real-time routing. The path computation involves using a formulated equation which characterizes communication intent between transmitter and receiver devices. The features selected for the clustering algorithm contribute towards optimizing the energy conservation effort. This paper also utilizes an evolutionary sleep scheduling technique which can be optionally used to further boost network efficiency. This technique combines the benefits of Particle Swarm Optimization (PSO) and Genetic Algorithm (GA). The proposed routing protocol has been simulated and compared with two existing routing protocols in terms of metrics such as number of active nodes, energy dynamics and network coverage. The simulation results prove that the proposed protocol outperforms LEACH and FCM.

\end{abstract}


% \fi
\IEEEpeerreviewmaketitle

\section{Introduction}
\IEEEPARstart {I}{nternet} of Things (IoT) \cite{GUBBI20131645} constitute a network of heterogeneous devices communicating and exchanging data amongst themselves to provide smarter services to users. The field of IoT has been witnessing increased research and development in several application areas. Smart home appliances and infrastructure, smart security and surveillance, smart road traffic management and medical emergency response systems are a few examples of IoT network's use cases. IoT networks are enormous in scale and complexity, and comprise objects like Radio Frequency Identification (RFID) tags, mobile phones and sensing devices to obtain data from the environment. Such devices, also referred to as sensor nodes, have low compute capability and limited battery life. The existing routing protocols for Wireless Sensor Networks (WSNs) \cite{1368893} are complex in nature and demand a considerable use of processing power and memory which are scarce resources in the devices comprising an IoT network. There is hence, a need for simpler protocols that are able to efficaciously conserve the energy of the devices. For an energy efficient network, and especially one that involves cooperative devices, it becomes important to distribute the processing load of routing evenly amongst all devices in the network. This ensures that a larger number of devices will remain operational for extended periods of time. On the contrary, if this processing load attributed to routing, is subjected to only a subset of the devices in the network, then this subset of devices would consume their energies at a faster rate and hence run out of battery sooner than the rest of the network. The proposed routing protocol in this paper aims to solve this problem by distributing the routing effort which in turn guarantees increased operational time of the network. The working of the protocol is detailed in later sections. Moreover, it must be noted that the term node and device both refer to the sensor nodes in the IoT network and have been interchangeably used in the text.

Apart from the sensor nodes, IoT networks also utilize base stations which are centers for data processing and storage. Base stations are more powerful as compared to the sensor nodes and hence come at a higher price. These are used in order to access network data and analyze it. There can be several models of IoT networks depending on the number of base stations used. For simulation of the proposed protocol, only a single base station is used as an end destination for the network data. The aim of the proposed protocol thus translates into finding an optimal energy-efficient path from the sensor nodes to the base station. Moreover, the protocol has to effectively minimize the energy expenditure of devices in the network and increase the network's operational lifetime. 

There exist several mechanisms which can supplement the routing protocol in a network and help conserve energy \cite{ANASTASI2009537}. One such popular approach is sleep scheduling. Sleep scheduling is widely employed in WSNs in order to avoid energy wastage which is caused by idle devices. Devices are chosen according to a predefined metric and are put to sleep (powered off) for fixed intervals of time through this mechanism. There exist several sleep scheduling algorithms \cite{kumar2011survey} that can be employed in an IoT-based WSN. Of these, approaches based on evolutionary algorithms have given promising results, and in particular \cite{6270782}, which introduces a Particle Swarm Optimization inspired sleep scheduling technique for WSNs, has been very successful. In this paper we utilize an amalgamation of two well known evolutionary techniques for the purpose of sleep scheduling: Genetic Algorithm (GA) \cite{davis1991handbook} and Particle Swarm Optimization (PSO) \cite{kennedy2011particle} algorithm. This sleep scheduling mechanism is referred to as Genetic Swarm Optimization (GSO) in this paper. The working of GSO is explained in later sections.  

To summarize, the contributions of this paper are as follows:

\begin{itemize}
\item Formulation of green routing in IoT as an energy optimization problem
\item Devising a minimum energy (MINEN) routing protocol as a solution to the aforementioned optimization problem
\item Supplementing the minimum energy routing protocol with a sleep scheduling mechanism, GSO, to enhance the energy conservation effort
\item  Performance comparison of the proposed protocol with two widely employed routing protocols as well as multiple sleep scheduling techniques
\end{itemize}

The rest of the paper is organized as follows. The problem statement is defined in Section II. Section III describes related work in green routing and deployment protocols for IoT-based WSNs. Section IV details the working of the proposed routing protocol. Section V presents the simulation performed and the results obtained. The scope for future work and the conclusion are present in Section VI.
%\begin{figure}
%\includegraphics[scale=0.3]{Anshuman.png} 
%\caption{}
%\end{figure}

\section{Problem Statement}

Sensor nodes in an IoT network generate data at a high and rapid rate. They communicate this data to the base station for processing purposes. They can also communicate with other sensor nodes for the purpose of eventually sending their data to the base station. Energy is spent both in transmission and reception of this message. The amount of energy used in one successful message transmission depends on several variables, such as the distance of the node from the base station, length of the message to be delivered, cost of power amplification before message transmission, and operational energy cost incurred by the transmitter and receiver hardware. Network communication links should be modeled in a manner which combines these variable costs to give a minimized additive output. The aim of a \emph{green} routing protocol should be to compute a path among nodes from the sender to the base station, which minimizes this cost.

Moreover, not every node is actively involved in the process of message generation and transmission at all times. Devices experiencing periods of such inactivity, wastefully consume energy waiting for interrupts and events to occur. Such wasteful usage of energy should be mitigated to contribute to energy cost optimization. This can be achieved by putting these devices to \emph{sleep}.

Nodes can also be clustered into groups and elected cluster `heads' can aggregate messages for all the nodes in their group/cluster and collectively transmit them to the base station. This has been shown to reduce energy consumption in the network. 

%In pursuit of energy cost minimization a routing protocol might divert network traffic to intermediate links with lesser costs. This would lead to rapid depletion of energy of select devices connected through such routes as these devices would be incessantly involved in routing related processing tasks. An ideal routing protocol must also incorporate traffic load balancing along with energy conservation in order to support high performance and efficiency of IoT networks.

A solution to the aforementioned problem(s) must meet the following objectives:
\begin{itemize}
\item The energy spent in transmission and reception of messages amongst the sensor nodes should be minimized
\item Sensor nodes should be clustered together in an efficient manner so that their messages are transmitted jointly and the network lifetime is increased
\item Effective sleep scheduling measures may be incorporated into the algorithm to further augment the energy conservation process
\end{itemize}

We detail our proposed approach based on the above objectives in Section IV.

\section{Related Works}

Recently, many different protocols have been proposed for wireless networks that ameliorate security issues, improve network reliability and boost energy efficiency \cite{doi:10.1002/dac.3487,7926114,DIPIETRO20141}. Machine learning and game theory have also been employed in computing routing solutions for wireless networks \cite{7926113}, \cite{VASHISHTH2019138}. Due to the energy considerations associated with most battery-based wireless networks, prolonging network lifetime is a key issue. This has amounted to an increased research interest in realization of a \emph{green} wireless IoT network. The authors in \cite{shaikh2017enabling, zhu2015green} discuss the need for green IoT and the various software and hardware based technologies required to enable its realization. Energy efficient inter-node communication and improved routing techniques have been identified as the issues that need to be addressed to facilitate large-scale adoption of green IoT. 

There exist several routing protocols and network deployment schemes that have been designed for WSNs. In \cite{1368893} routing protocols for WSNs have been categorized into 3 different groups: Flat routing, location-based routing and hierarchical routing. The paper \cite{1368893} presents an extensive survey of routing protocols categorized under each of these three groups. Flat routing protocols \cite{Savvides:2001:DFL:381677.381693, 1024422,Braginsky:2002:RRA:570738.570742} are data centric, which implies that there are no stringent constraints on the origin of the data. Sensor nodes in such protocols collaborate to perform sensing tasks and data is queried by the base station from different geographical regions. Location based routing protocols like \cite{869214, 878532} address sensor nodes by their locations. This is made possible through various methods including relative distance estimation using signal strengths or information exchange and GPS based location tracking. The last category of protocols is relevant to the problem of energy conservation. Hierarchical routing \cite{878533,956276,Heinzelman:1999:API:313451.313529,heinzelman2000energy} is the category of protocols which efficiently divide network's routing responsibilities on the basis of device capabilities. High energy nodes are assigned with the task of data processing and data transfer, whereas low energy devices are tasked with sensing the environment. Cluster based routing techniques are also grouped under hierarchical routing as the specialized data processing and transmission tasks are assigned to the selected cluster heads. The proposed routing mechanism in this paper clusters devices on the basis of a number of relevant features, such as the distance of the node from the base station, residual battery levels of the devices, length of messages generated and amount of data sensed by the sensor nodes. 

LEACH \cite{heinzelman2000energy} is a hierarchical routing protocol which involves clustering of devices for transmitting data to the base station. Cluster heads are responsible for data aggregation, data processing and data communication. The protocol involves randomized cluster head rotation for distribution of routing load among multiple sensors. Data aggregation and fusion is also employed to reduce the size of data to be transmitted. LEACH is one of the most widely adopted energy-efficient routing algorithms for sensor networks and we treat it as a baseline for our approach. From a historical perspective other energy-efficient approaches have also been proposed: Minimum Transmission Energy protocol (MTE) \cite{ettus1998system} routes data through the nearest neighbor nodes to minimize transmission energy. However, when the network uses MTE, nodes nearest to the base station are overburdened with routing related processing load and hence run out of battery very fast. Through collaborative routing, such unequal distribution of routing effort has been mitigated in several protocols including LEACH and FCM \cite{hoang2010fuzzy}. % delete explanation (MTE and DC ) if Anshuman does not approve

LEACH suffers from some drawbacks as well. First, the random rotation of cluster heads may lead to several sub optimal periods of communication. This is because inadequate nodes could be elected as cluster heads in these periods. Second, the distance of all cluster heads in the network is not uniform, hence some cluster heads would have to transmit data at longer distances than others. Both of these drawbacks are accounted for in our proposed protocol where cluster heads are elected on the basis of residual battery levels of the devices. Moreover, data transmission to the base station is undertaken as a collaborative effort by all the cluster heads in the network. An energy efficient route comprising of all the cluster heads is computed for the eventual transmission of data.

FCM \cite{hoang2010fuzzy} is another clustering based routing protocol. FCM proposes clustering on the basis of minimization of the euclidean distance of devices from the centers of the clusters they belong to. This is done to create an energy balance amongst sensor nodes in the cluster. Cluster heads in FCM are elected on the basis of residual energy levels. Cluster heads aggregate data from all the devices in the network and transmit it to the base station. FCM does not take into account the data generation capacity of devices in formation of clusters. Hence if a majority of devices which generate lengthy messages and actively sense larger amounts of data than others are clubbed into one cluster, then the assigned cluster head of this group would incur heavier energy expenditure than the others in the network. Eventually, this cluster would run out of battery earlier than others. This drawback is compensated for in our protocol as well, as we consider three device characteristics for cluster formation: distance of nodes from the base station, length of messages generated by devices and amount of data sensed by the sensor nodes in one communication epoch.

Our proposed routing protocol is simulated and compared with LEACH and FCM in section V. It will be seen that our protocol outperforms both LEACH and FCM on the basis of several metrics. 

Research has also been put into development of network deployment schemes in order to achieve green IoT. One such scheme is proposed in \cite{huang2014novel}. It is a static network deployment scheme which utilizes two kinds of nodes, sensor nodes and relay nodes. Relay nodes are considered to be more computationally powerful than sensor nodes and are used to manage clusters of sensor nodes. The network is constructed into a three tiered hierarchy where sensor nodes make up the lowest level, relay nodes the intermediate level and base stations comprise the top most level. Huang et al have modeled communication constraints amongst nodes in the network according to the hierarchy level of the node. The authors represent the energy and cost optimization problem as a routing problem. The Steiner tree algorithm is used to solve this optimization problem of computing the number and position of relays in the network. The shortcoming of this approach is that it is a static deployment scheme and places many restrictions on how the network will have to be designed. In our approach this problem is ameliorated, as there is no static hierarchy between different nodes, the overall approach is simplistic in nature, and can be adopted to conserve energy in any kind of sensor network application.

As mentioned before, sleep scheduling is an efficient mechanism to supplement energy conservation in an IoT network. In \cite{6270782} an energy balancing sleep scheduling mechanism is proposed for WSNs, which is based on Particle Swarm Optimization. In this paper, we utilize a sleep scheduling mechanism based on the Genetic Algorithm (GA) and Particle Swarm Optimization (PSO) evolutionary approaches and refer to it as Genetic Swarm Optimization (GSO). The performance of the proposed protocol is compared to three other sleep scheduling techniques: GA, PSO and ECCA \cite{JIA20091756}. The results from the simulations show that the proposed protocol gives best results when supplemented with GSO. We have analyzed the performance of the proposed Minimum Energy algorithm (MINEN) with and without the use of sleep scheduling, as sleep scheduling can lead to more complexity which may be undesirable in some networks. From here on in the text, MINEN will signify MINEN without sleep scheduling. In case sleep scheduling using GSO is also employed, it will be explicitly referred to as `MINEN with GSO'.

\section{Proposed Protocol}

The flow of the proposed minimum energy (MINEN) routing protocol is depicted in Figure 1. The major steps of the protocol can be summarized as follows:

\begin{itemize}
\item Running sleep scheduling to identify nodes not participating in the current epoch. This step is optional as the performance of the protocol is also analyzed without the use of sleep scheduling. 
\item Cluster formation and cluster head election
\item Construction of DAG connecting all cluster heads and calculation of edge weights
\item Running Dijkstra to identify the minimum cost path to the base station for the current epoch
\end{itemize}

\begin{figure}
\centering
\includegraphics[scale=0.40]{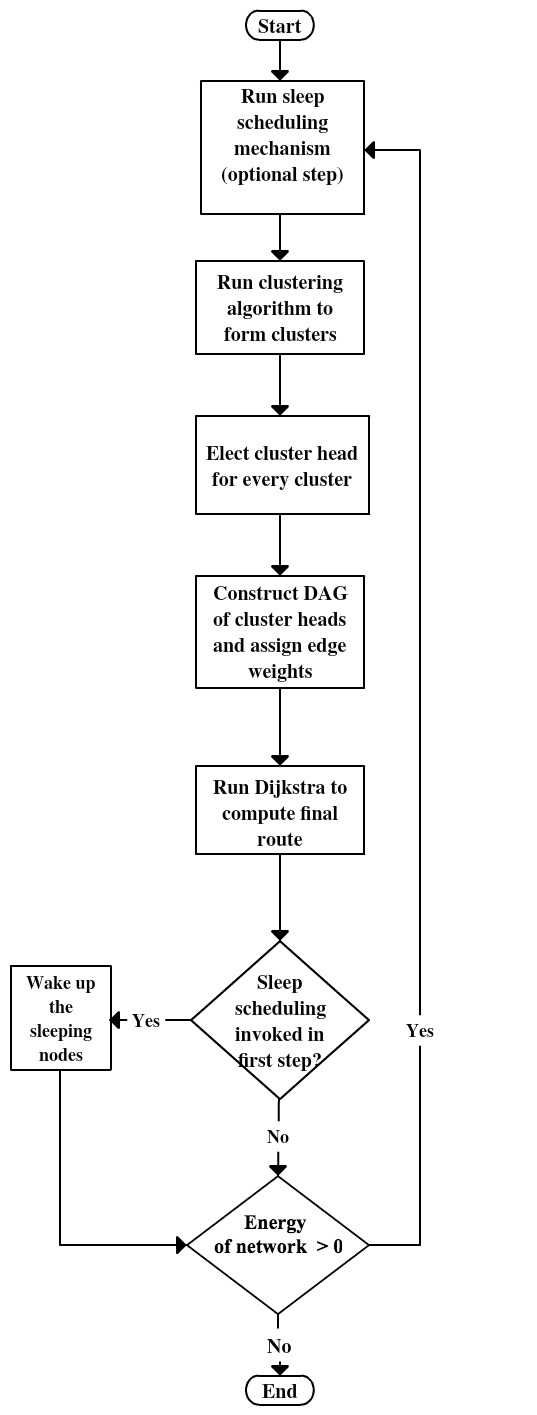}
\caption{Flow of the algorithm}
\end{figure}

The assumptions made by the protocol are:
\begin{itemize}
\item All devices begin with the same level of energy
\item There is only one base station located at a static position in the IoT network
\item Base station is assumed to be supplied with infinite amount of energy, i.e. a base station is not at the risk of shutting down due to lack of energy
\item A \emph{round} of communication is assumed to be the period of time between election of new cluster heads in the network and successful transmission of messages from all cluster heads to the base station
%right about communication round here
\end{itemize}

We now explicitly detail each step of the algorithm shown in Figure 1 in the following subsections:

\subsection{Sleep Scheduling}

Sleep scheduling is used to identify idle devices which should be powered-off at the start of every communication round, to save energy. As mentioned before, in this paper we utilize GSO for sleep scheduling, which is a combination of PSO and GA. These evolutionary optimization techniques typically progress through the following sequence of steps: 

\begin{enumerate}
\item Randomly generate an initial population of solutions
\item Calculate a fitness value for each solution. This fitness value must be representative of the closeness of the current solution to the optimum solution
\item Reconstruction and modification of the population of solutions on the basis of their fitness values
\item Repeating the process from step 2 until requirements for the ideal solution have been met
\end{enumerate}

A typical solution is considered to be a boolean array of the size of the number of nodes in the network. Each index of the array contains a boolean value signifying whether the device at that index should be put to sleep (true) or not (false). A set of such arrays is referred to as a population of solutions. This population is randomly initialized in the beginning and is altered using a set of operators until the final solution is achieved. The difference between PSO and GA lies in the way this population of solutions is altered. 

GA utilizes two set of operators in the process of modification and alteration of the solution population called mutation and crossover. The mutation operation makes adjustments to a particular solution candidate. To explain this in terms of data structures, a mutation operator masks randomly chosen elements of the array. The crossover operator on the other hand creates a new solution array from the combination of two previously existing and randomly selected solutions, referred to as parent solutions. This new array is formed by taking some element values from one parent solution array and the rest of the elements from the other. Every element in the new solution array can take two possible values corresponding to the elements at the same index in the parent solutions. A random probabilistic function is used to make this decision. GA sleep scheduling technique chooses two solution arrays, performs mutation operation on each array and then carries out the crossover operation to generate the next generation of solutions. PSO on the other hand adjusts the solution population on the basis of two parameters, the local best fitness value of the solution array (best fitness of a subset of solutions of the solution space) and the globally best fitness value identified so far, in the entire solution space. 

GSO combines the characteristics of both PSO and GA. GSO also uses the mutation and crossover operations. However, the way solution arrays are subjected to these operations is different than GA and PSO. Before explaining the modification step of the GSO algorithm, it is important to understand the way the solution array's fitness values are computed. The fitness value of every solution set is calculated using the following formula:

\begin{equation}
\textrm{Fitness value} = \alpha . term_1 + \beta . term_2
\end{equation}

where,
\begin{equation}
term_1 = 1 - \frac{partial\_weighted\_energy}{total\_weighted\_energy}
\end{equation}

\begin{equation}
term_2 = 1 - \frac{partial\_coverage}{total\_coverage}
\end{equation}

Here the $partial\_weighted\_energy$ corresponds to the sum of the energies of all the awake devices in the particular solution set and $total\_weighted\_energy$ signifies the collective sum of the energies of all the devices of the network. The $partial\_coverage$ corresponds to the network coverage provided by all the awake devices in the particular solution whereas $total\_coverage$ corresponds to the coverage provided by all the devices of the network. $\alpha$ and $\beta$ are used as the weights of the fitness function where $\alpha$ is used to optimize the energy lifetime and $\beta$ is used to optimize the network coverage.

The reconstruction and modification of the solution space follows the pseudo code described in Algorithm 1. This basically makes up the third step of a typical evolutionary sleep scheduling technique, as stated before. GSO is made to execute (steps 2 to 4) for a predefined number of maximum iterations - \emph{M}, after which the variable $global\_best$ will store the final sleep schedule to be used in MINEN. 

\begin{algorithm}
\caption{GSO Optimization step}
 \begin{algorithmic}[1]
 \State \textbf{Input: vector} solution\_population
 \State max\_solutions = length(solution\_population)
  \For{it = 0 to max\_solutions}
   \State current\_solution = solution\_population[it]
   \State \textbf{Mutate}(current\_solution)
   \State $crossover\_rate_1$ = 1 - it/max\_solutions
   \State $crossover\_rate_2$ = it/max\_solutions
   \If{random\_generator $<$ $crossover\_rate_1$}
   \State \textbf{Crossover}(current\_solution, local\_best)
   \ElsIf{random\_generator $<$ $crossover\_rate_2$}
   \State \textbf{Crossover}(current\_solution, global\_best)
   \EndIf
   \If{current\_solution.fitness $>$ local\_best.fitness}
   \State local\_best = current\_solution
   \EndIf
  \EndFor
 \end{algorithmic}
\end{algorithm}

The use of GSO is preferred over both PSO and GA. In GA the entire solution population moves together towards the optimal solution because every solution in the successive generation is influenced by two randomly chosen solutions in the present iteration. When constraining one of these parent solutions to be a local or global best solution, we increase the pace with which GSO converges to the optimal solution. This constraint also helps guide the evolution to only look for the best available solution.

\subsection{Clustering:}

The next step of the MINEN protocol involves dividing the IoT network of devices into clusters. We employ two clustering algorithms as separate experiments in this regard: K-means clustering and Gaussian Mixture Model (GMM) clustering. The performance of MINEN algorithm, when using both of these clustering techniques is later compared in Section IV, to see which clustering approach performs better. 

Clusters are formed on the basis of the similarity of certain chosen features of devices. In our approach the features used for clustering are the distance of the node from the base station, the length of messages generated by devices and the amount of data sensed by a device in one round of communication. K-means clustering technique divides the network of devices into 'K' clusters. During the clustering process each device is added to the cluster having the nearest mean value to the current device's features. Gaussian Mixture Models (GMM) unlike K-means, perform clustering under the assumption that the feature sets are normally distributed in space. The probability distribution is thus assumed to be as follows:
\begin{equation}
p(X) = \sum_{i=1}^{N}\pi\textsubscript{i}\mathcal{N}(X | \mu\textsubscript{i}, \Sigma) 
\end{equation}

Here $\pi\textsubscript{i}$ refers to the ratio of the data points present in the $i_{th}$ Gaussian cluster component and is known as the mixture coefficient. It's value ranges from between 0 and 1 depending on the cluster, and the total sum of all mixture coefficients will be equal to 1. $\mu_{i}$ is the mean of the entire $i_{th}$ Gaussian cluster and $\Sigma$ is the covariance. Also $N$ refers to the total number of clusters chosen. Under this assumption of the probability distribution, clustering is undertaken using a log-likelihood maximization algorithm. Therefore, if this assumption holds, GMM will outperform K-means clustering. Conversely, if the underlying assumption is false, then K-means will fare better and give more suitable clusters. Recently, Gaussian Mixture Models have been used with great success to solve various problems in communication and networking \cite{bahrololum2008anomaly,chhabraclassifying,sheng2005distributed}. 

\subsection{Cluster head election:}

After the formation of clusters, each cluster elects a cluster head. The cluster head is responsible for aggregating messages from all the devices in its cluster and forwarding them to the base station. Cluster heads hence do the \emph{heavy lifting} for their clusters in terms of processing and energy expenditure, and are involved in the routing path formation to be discussed later. However, this leads to cluster heads running out of energy rather fast. To avoid cluster heads from becoming inoperable due to loss of battery power, MINEN re-elects a new cluster head after a round has passed. In the beginning of every round, cluster heads are elected on the basis of the residual battery levels. The device with the maximum residual energy in a given cluster becomes the cluster head for that subsequent round. This rotation of cluster heads leads to an even distribution of routing effort amongst devices and hence increases the network's operational lifetime. Since in the first round every device will have the same level of energy, cluster heads in this round are chosen randomly.

After election of cluster heads, a Directed Acyclic Graph (DAG) connecting all of these cluster heads is created. Before describing the construction of this DAG, we lay down the energy model adopted by us to calculate energy expended by the network devices.

\subsection{Energy Model}

It is necessary to understand the energy considerations amongst devices in an IoT network. Energy is utilized in both transmission and reception of a message. The following variable definitions are used to derive energy equations for the energy model:

\begin{itemize}
\item $E_{r(ij)}$ - energy expended in reception of a message between devices i and j. 
\item $E_{t(ij)}$ - energy spent in transmission of a message between devices i and j.
\item $d_{ij}$ - distance between devices i and j.
\item $d_{o}$ - threshold distance
\item $l_{ij}$ - length of a message exchanged between devices i and j.
\item $E_{elec}$ - energy consumed to operate the transmitter or receiver circuit.
\item $\epsilon_{MP}$ - energy dissipated at power amplifier considering a multi-path fading channel.
\item $\epsilon_{FS}$ - energy dissipated at power amplifier considering a line-of-sight free-space channel.
\item $R_{ij}$ - rate of data transmission between devices i and j.
\item $e_{r(ij)}$ - energy of message reception per unit time between devices i and j.
\item $e_{t(ij)}$ - energy of message transmission per unit time between devices i and j.
\item $e_i$ - current energy of device i.
\item $e_j$ - current energy of device j.
\item $I$ - initial value of energy for all devices.
\item $E_{sf(ij)}$ - energy spent so far of devices i and j.
\item $w_1$, $w_2$, $w_3$ - weights assigned to edge weight components.
\end{itemize}

According to the Friis free space model \cite{229847} we have,

\begin{equation}
E_{t(ij)} = (E_{elec} + \epsilon_{FS} . d_{ij}^2) . l_{ij} \quad \textrm{for} \quad d_{ij} < d_o
\end{equation}

\begin{equation}
E_{t(ij)} = (E_{elec} + \epsilon_{MP} . d_{ij}^4) . l_{ij} \quad \textrm{for} \quad d_{ij} >= d_o
\end{equation}

\begin{equation}
E_{r(ij)} = E_{elec}.l_{ij}
\end{equation}

As can be seen from the equations, the dependence of transmission energy on distance increases by a power of 2 after a threshold distance $d_{o}$. Length of data sent per unit of time $(t)$ is the rate of data transmission between devices $i$ and $j$. Hence,

\begin{equation}
R_{ij} = l_{ij}/t
\end{equation}

Using equation 8, equations of energy of transmission and reception of messages per unit time amongst devices i and j are written as:

\begin{equation}
e_{t(ij)} = (E_{elec} + \epsilon_{FS} . d_{ij}^2) . R_{ij} \quad \textrm{for} \quad d_{ij} < d_o
\end{equation}

\begin{equation}
e_{t(ij)} = (E_{elec} + \epsilon_{MP} . d_{ij}^4) . R_{ij} \quad \textrm{for} \quad d_{ij} >= d_o
\end{equation}

\begin{equation}
e_{r(ij)} = E_{elec}.R_{ij}
\end{equation}

\subsection{Graph construction and route formation}

After cluster formation and cluster head election, a directed acyclic graph (DAG) connecting all the cluster heads and the base station is created. At this point, the cluster heads have already aggregated the data from all the sensor nodes in their cluster. Thus, equations (9, 10, 11) for the cluster heads reflect the transmission and reception of this data as messages. Each edge connecting two cluster heads or a cluster head and a base station are assigned certain edge costs/weights on which Djikstra's algorithm is applied to find the minimum cost path. These weights incorporate the energy of message transmission and reception (equations (9, 10, 11)) as well as the energy spent so far $(E_{sf})$, of the devices (cluster heads) on a directed edge. Mathematically the $E_{sf}$ corresponds to,

\begin{equation}
E_{sf(ij)} = (I - e_i) + (I - e_j)
\end{equation}

$E_{sf}$ has been included in the edge weight in order to introduce load balancing in terms of energy, across devices in the network. This can be explained using the following example. Consider two intermediate route links, one of which is to be included in the least energy cost path to the base station where each device has an initial energy ($I$) of 5J. These links have an additive reception and transmission energy cost (that is, $e_{r(ij)}+e_{t(ij)}$) of 2J and 10J, respectively. However, the additive residual energy levels (that is, $e_{i} + e_{j}$) of these device link pairs correspond to 1J and 10J respectively. Knowing that Djikstra's algorithm choses the least cost path to the base station, if the cost does not include $E_{sf}$ of devices then the 2J link will always be opted for over the 10J link despite the residual energy of the device pairs in the first link being much lower. However, if $E_{sf}$ is included in an additive manner, the edge costs would actually be 11J and 10J for these links, respectively. Therefore, Djikstra's algorithm would actually select the latter link over the former link which is the kind of behaviour we would like the proposed algorithm to exhibit. Mathematically the value of costs (edge weights) assigned to edges on the graph is a weighted sum of three components,

\begin{equation}
e_{ij} = w_1.e_{r(ij)}+ w_2.e_{t(ij)} + w_3.E_{sf(ij)}
\end{equation}

A weighted sum is taken in order to fine tune the impact of every component on the performance of the protocol. This can be required for some networks that might be transmission cost heavy and others that might have a higher energy cost for reception of messages. These weights allow the designer to give specialized preference depending on the type of network. However, setting all these weight values ($w_1$, $w_2$, $w_3$) to 1 is generally the best setting for the algorithm. All simulations carried out for MINEN take $w_1 = w_2 = w_3 = 1$. 

It should be noted that costs assigned to edges connecting base stations with the cluster heads ignores the reception energy and the $‘E_{sf}’$ of the base station. Mathematically,

\begin{equation}
e_{ib} = w_2.e_{t(ib)} + w_3.E_{sf(i)}
\end{equation}

\begin{equation}
E_{sf(i)} = I - e_i
\end{equation}

Here $b$ refers to the base station. This is because of the assumption that the base station never runs out of energy. Hence the cost of energy reception by the base station is not required to be minimized. 

Once the graph is constructed we run Dijkstra algorithm to find the minimum energy path from every cluster head to the base station. Messages are then passed through this route for the duration of one round.

\subsection{Flow of the algorithm}
The steps of the proposed protocol MINEN with GSO (or just MINEN) are formally summarized using Algorithm 2. Step 1 runs the sleep scheduling algorithm if required. Step 2 formulates clusters in the network. The sequence of steps from 3 to 7 help elect a cluster head by selecting the device with the maximum residual energy corresponding to each cluster in the network. Steps 8 and 9 construct a DAG connecting all cluster heads and Dijkstra's algorithm is then used to return the minimum cost routing path for one round. If sleep scheduling has been invoked in the beginning, then steps 11 and 12 help wake up the relevant nodes at the end of every round. This algorithm continues until all the devices of the network run out of battery. Figure 2 graphically shows the working of MINEN in a sample network.

\begin{figure}
\centering
\includegraphics[scale=0.28]{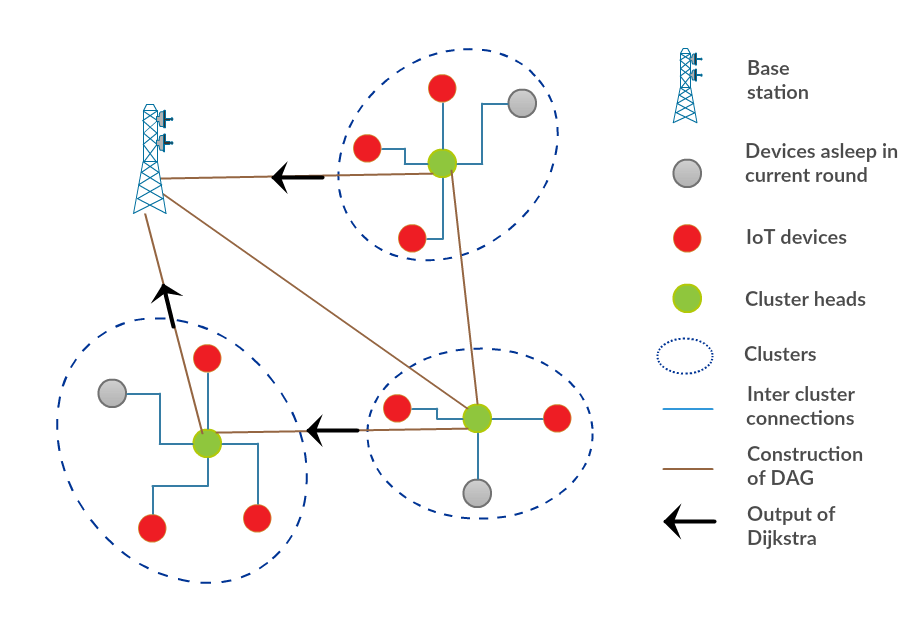}
\caption{MINEN routing on a sample IoT network}
\end{figure}

\begin{algorithm}
\caption{MINEN}
 \begin{algorithmic}[1]
 \State $\textbf{\emph{Run}} \textrm{ sleep scheduling algorithm \emph{GSO} (optional)}$
 \State $\textbf{\emph{Run}} \textrm{ clustering algorithm to create cluster set C}$
 \For{$c \in C$}
  \For{$i \in c.devices$}
   \If{$d.current\_energy < i.current\_energy$}
   \State $d = i$
   \EndIf
  \EndFor
   \State $\textbf{\emph{Set }} c.cluster\_head = d$
 \EndFor
\State \textbf{\emph{Create}} graph G connecting all cluster heads
\State \textbf{\emph{Run}} \emph{Dijkstra} on G
\State \textbf{\emph{Return}} routing path
\If{\emph{Sleep scheduling is invoked in step 1}}
\State \textbf{\emph{Wake}}\textrm{ up all the sleeping devices in the network}
\EndIf
\If{\textrm{Number of active devices$ > 0$}}
	\State \textbf{\emph{goto}} step 1
\Else{\textrm{\emph{ exit}}}
\EndIf
 
 \end{algorithmic}
\end{algorithm}

\section{Simulations and Results}

This section discusses the results of the simulations done for demonstrating the efficacy of the proposed protocol. We have used the simulator implemented in \cite{wsn} for running our simulations. We simulated MINEN, LEACH and FCM for comparative analysis on the basis of :

\begin{figure}
\centering
\includegraphics[scale=0.46]{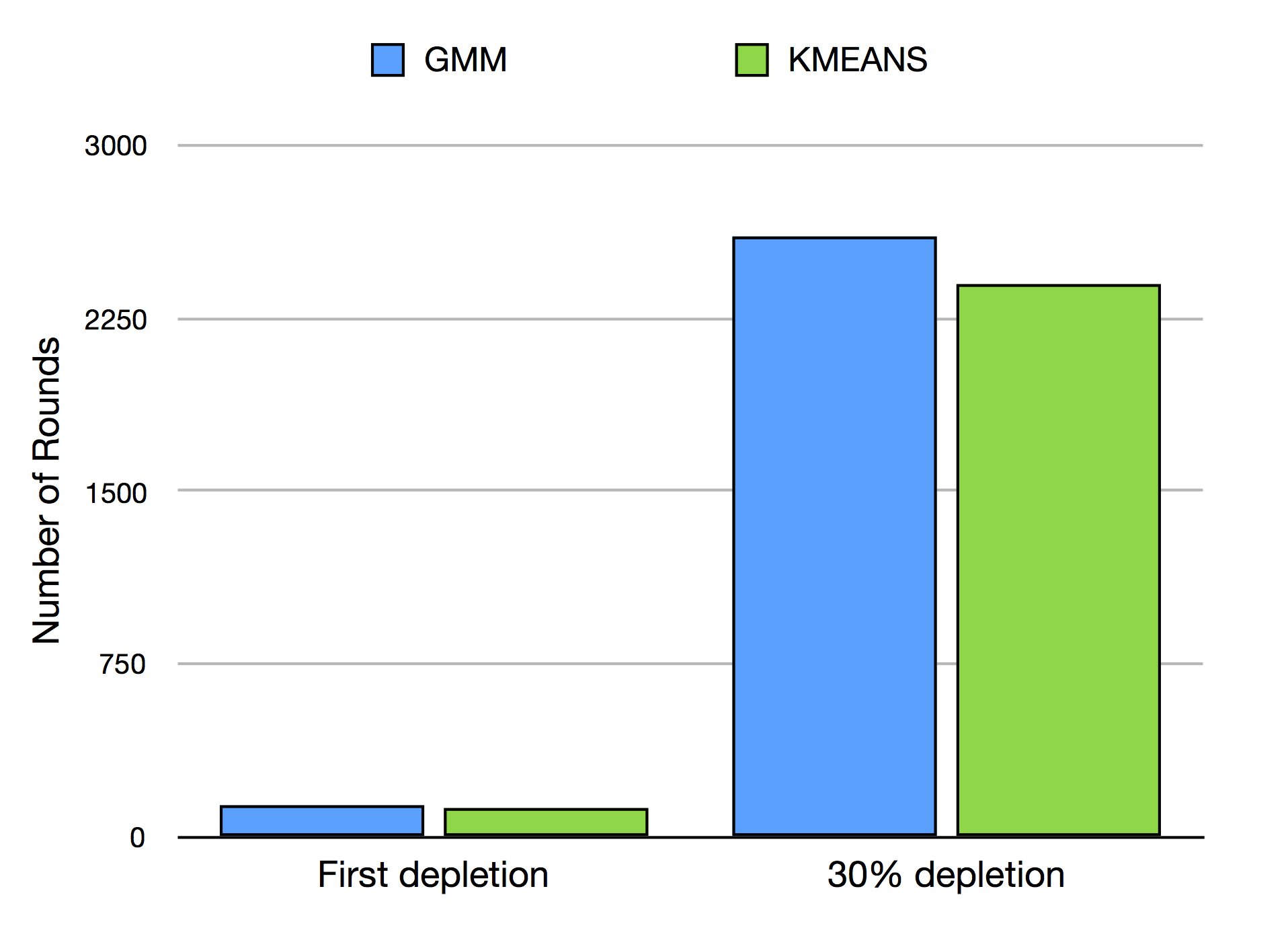}
\caption{Clustering algorithms comparison}
\end{figure}

\begin{itemize}
\item \emph{Number of alive nodes}: This evaluation observes the quantity of actively participating nodes through the progression of communication rounds. The higher the number of active devices, the more would be the productive capacity of the network.
\item \emph{Energy dynamics}: An energy versus number of rounds (time) comparison has been done to calculate the number of rounds for which the energy of a network lasts. 
\item \emph{Network Coverage}: This is used to analyze the geographical coverage of the operating devices of the network through the progression of time. It can be defined as the number of rounds for which devices are actively contributing to the IoT network's operation in a particular geographical region.
\end{itemize}

Table I displays the simulation parameter's initialization values that have been used in analysis of the results of the proposed protocol in this section. We have analyzed the performance of MINEN with and without the use of sleep scheduling. It will be seen that even without the additional energy conservation support of sleep scheduling, MINEN outperforms LEACH and FCM by considerable margins. Since MINEN, LEACH and FCM perform randomized clustering, different results can be obtained every time the algorithms are run. Therefore, the results shown in this section are the \emph{best} possible results obtained for MINEN. Results for FCM and LEACH are randomly obtained.

\begin{table}[h!]
\centering
\caption{Simulation parameters}
\begin{tabular}{||c c||}
\hline
Parameters & Values \\ [0.5ex]
\hline\hline
$w_1$ & 1 \\
$w_2$ & 1 \\
$w_3$ & 1 \\
$I$ & 2 J \\
Number of nodes & 300 \\
Simulation area & $250*250$ m\textsuperscript{2} \\
$E_{elec}$ & 50e-9 J/bit \\
$\epsilon_{MP}$ & 0.0013e-12 J/bit.m\textsuperscript{4}\\
$\epsilon_{FS}$ & 10e-12 J/bit.m\textsuperscript{2}\\
$d_o$ & $\sqrt[]{\epsilon_{FS}/\epsilon_{MP}}$ \\
$\alpha$ & 0.34 \\
$\beta$ & 0.33 \\ 
\emph{M} & 50 \\[1ex]
\hline
\end{tabular}
\end{table}

For analyzing the effect of clustering on the performance of the proposed routing protocol we simulate MINEN using both GMM and K-means. Figure 3 depicts this comparison. Some research \cite{huang2014novel} defines network lifetime as the time duration beginning from the start of the network execution to the depletion of first device in the network. As can be seen from figure 3 this value is almost the same when both GMM and K-means are used in MINEN. However the number of communication rounds after which 30\% of the nodes in the network have run out of energy is higher for GMM than for K-means. Thus GMM performs better. The rest of the simulation results have been computed using GMM as the clustering algorithm in MINEN. %confirm this with Anshuman.

\begin{figure}
\centering
\includegraphics[scale=0.433]{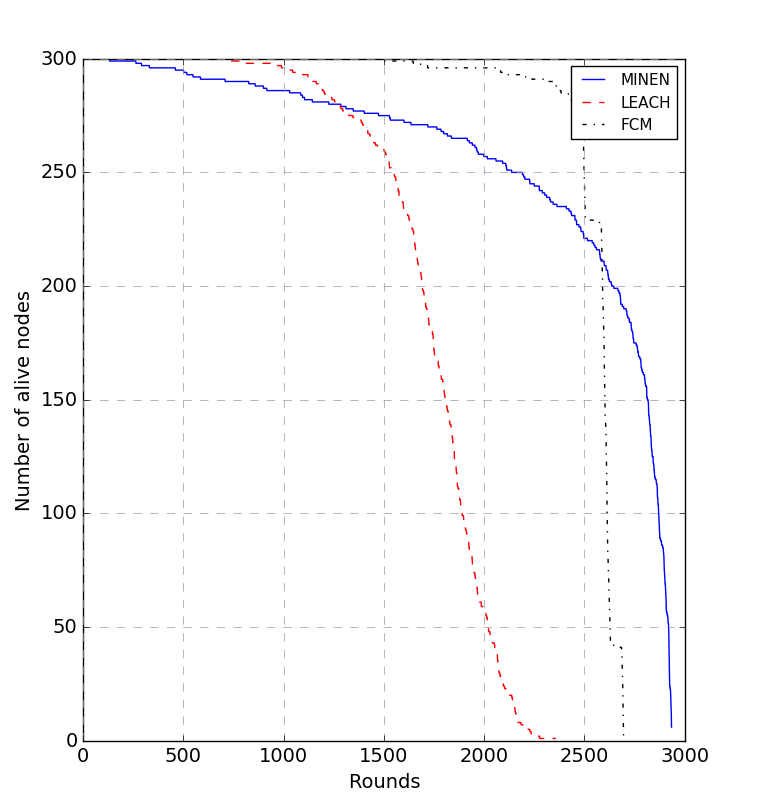}
\caption{}
\end{figure}

\begin{figure}
\centering
\includegraphics[scale=0.43]{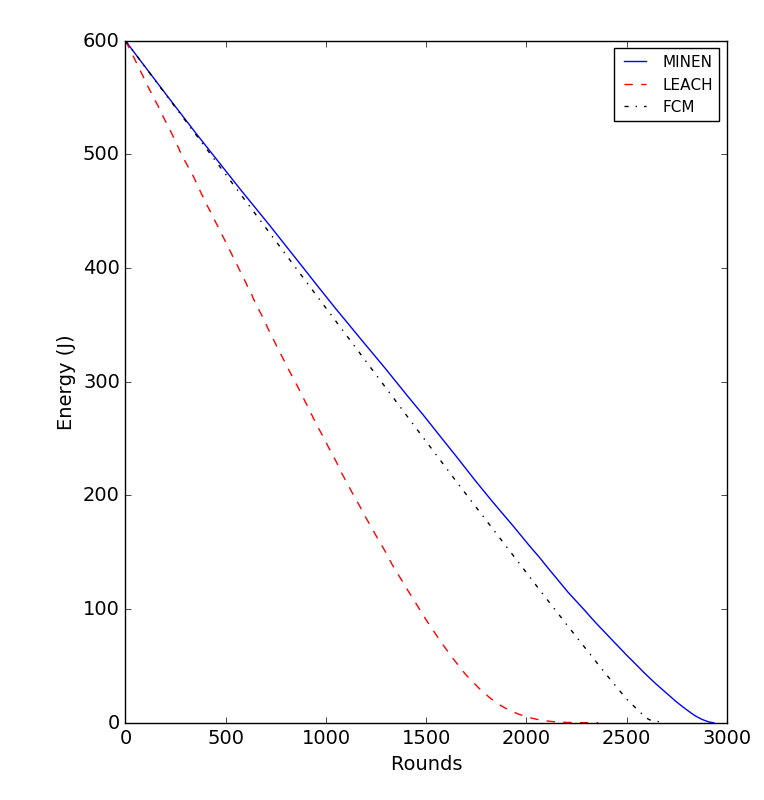}
\caption{}
\end{figure}

Figure 4 plots the \emph{number of alive nodes} in the IoT network against the number of rounds of communication (time) in the simulation. As can be seen from the graph, in a network of 300 devices, MINEN is able to keep 150 of these devices operational for up to 2800 rounds. LEACH and FCM on the other hand keep 150 devices active only up to 1700 and 2600 rounds respectively. Applying MINEN increases network application's operational time to a larger extent than LEACH and FCM. This signifies that MINEN introduces more efficient distribution of routing effort amongst devices. Figure 5 depicts the \emph{energy dynamics} of the IoT network. Energy of the network will drop down to 0 when all devices run out of battery. For MINEN this happens after approximately 3000 rounds of operation, LEACH keeps the network operational for roughly 2100 rounds and FCM accomplishes the same in about 2600 rounds. The slope of the graphs also depict that LEACH has the steepest energy depletion rate followed by FCM and MINEN respectively. Figure 6 shows the number of rounds after which 30\% of the network's devices run out of energy and become inoperable. For MINEN 30\% depletion occurs after about 2608 rounds. For LEACH and FCM these values are 1671 and 2589 respectively. Figures 4, 5 and 6 conclusively prove the better performance of MINEN over LEACH and FCM in terms of energy efficiency. This performance is due to a better choice of features used in clustering, better cluster head selection criteria and introduction of energy load balancing across different communication links in MINEN. All of these have been extensively discussed in the previous sections.  

\begin{figure}
\centering
\includegraphics[scale=0.46]{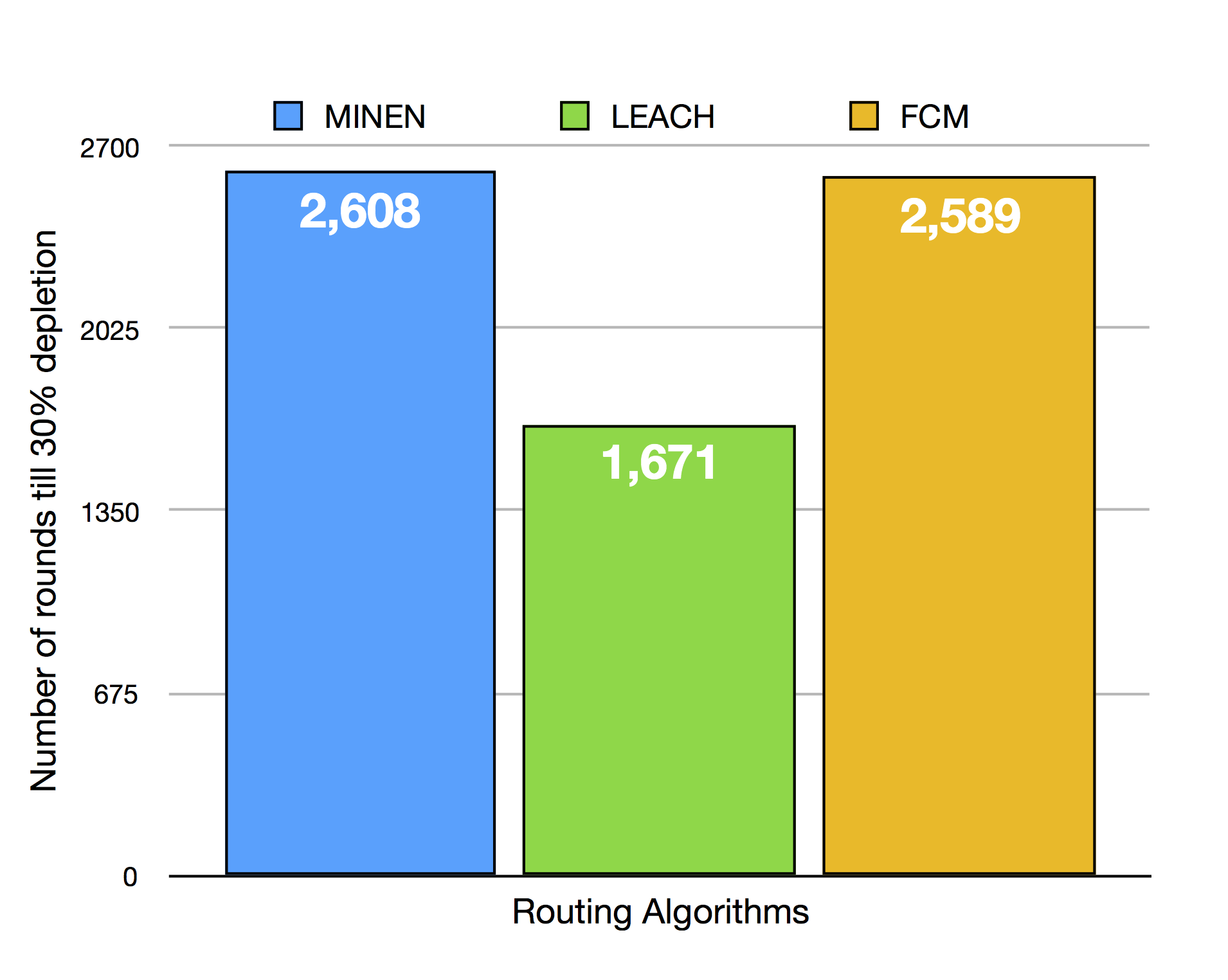}
\caption{}
\end{figure}

The next set of figures compare the three algorithms on the basis of \emph{network coverage}. Figure 7, 8, 9 and 10 showcase this comparison. A red cross in the middle of these graphs depicts the base station of the network. The figures depict the simulation area of the IoT network and are colored in different hues on the basis of how long devices in every section are active and operational. An efficient green routing algorithm should be able to keep active devices uniformly distributed across the network region. Moreover the operational time of devices should be long.  Figures 7 and 8 depict the network coverage provided by FCM and LEACH respectively. The network's geographical area is fragmented into three sub regions in figure 7. The sub region occupying the largest area (peach) in this figure corresponds to 2500 to 2600 rounds until full depletion. The lower left (green) subregion corresponds to an operational time of 2400-2500 rounds whereas the upper right (red) subregion corresponds to 2600-2700 rounds. Hence FCM provides uneven \emph{network coverage} across the simulation area. Figure 8 shows an even more non-uniform distribution of \emph{network coverage} provided by LEACH. The number of rounds until complete depletion of devices in LEACH lies in the range of 1000 to 2200. The sub regions where devices are active for longest period of time are small and scattered in the middle of the simulation region. As can be seen from the figure devices equidistant from the center provide the same network coverage. In other words, devices form somewhat concentric circular subregions, where the subregion with larger radius last for lesser number of rounds. As explained before, this happens because cluster heads which are at a farther distance from the base station expend more energy in communicating messages to the base station in LEACH.  

\begin{figure}
\centering
\includegraphics[scale=0.45]{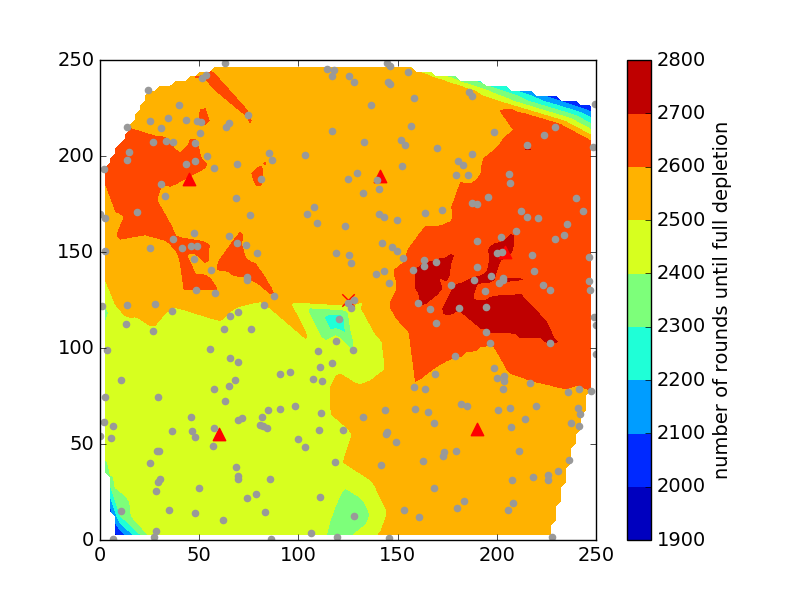}
\caption{FCM network coverage}
\end{figure}

\begin{figure}
\centering
\includegraphics[scale=0.45]{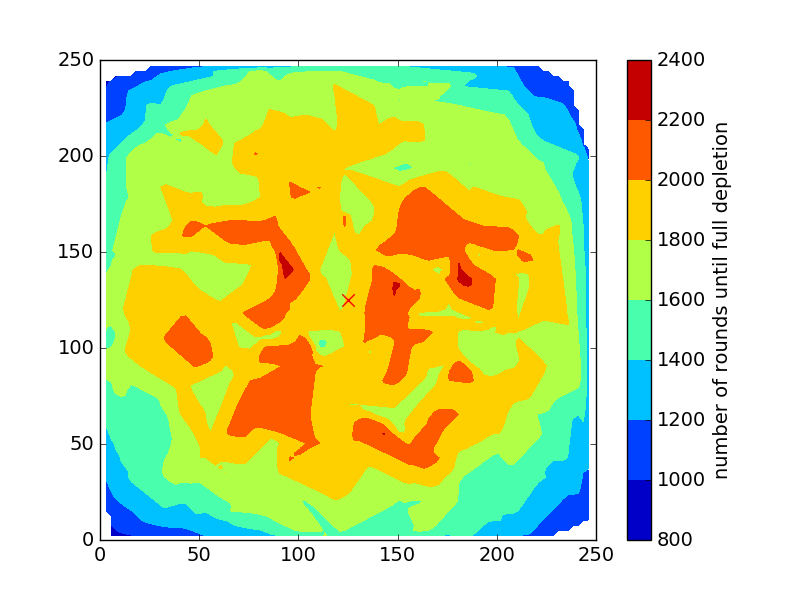}
\caption{LEACH network coverage}
\end{figure}

\begin{figure}
\centering
\includegraphics[scale=0.45]{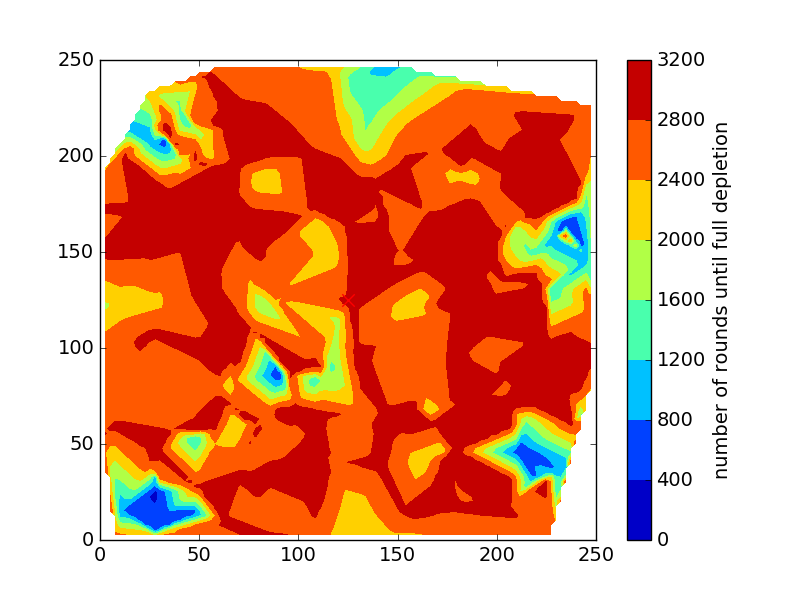}
\caption{MINEN network coverage}
\end{figure}

\begin{figure}
\centering
\includegraphics[scale=0.45]{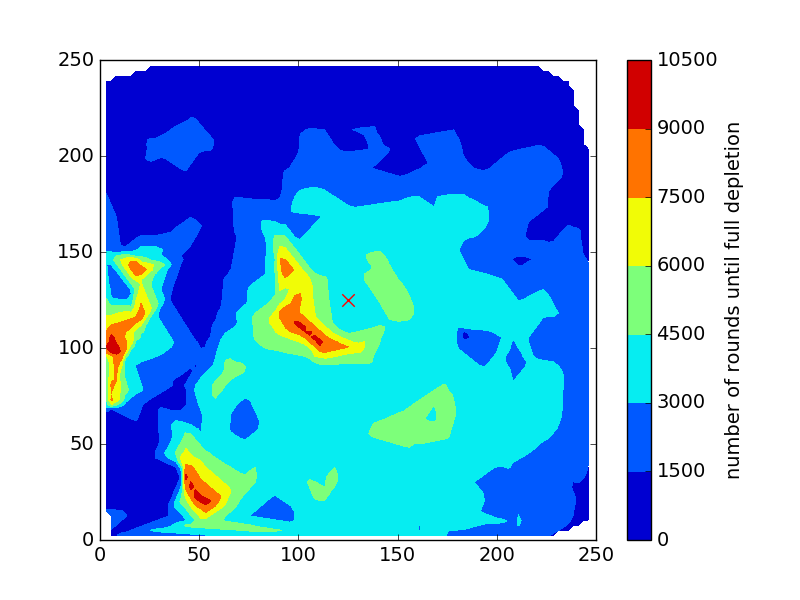}
\caption{MINEN with GSO network coverage}
\end{figure}

\begin{figure}
\centering
\includegraphics[scale=0.4]{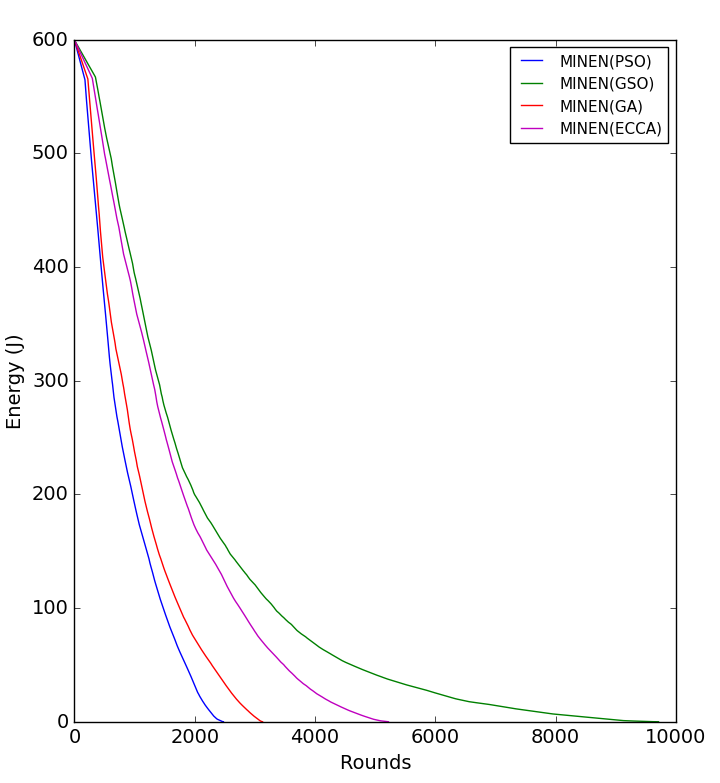}
\caption{}
\end{figure}

\begin{figure}
\centering
\includegraphics[scale=0.4]{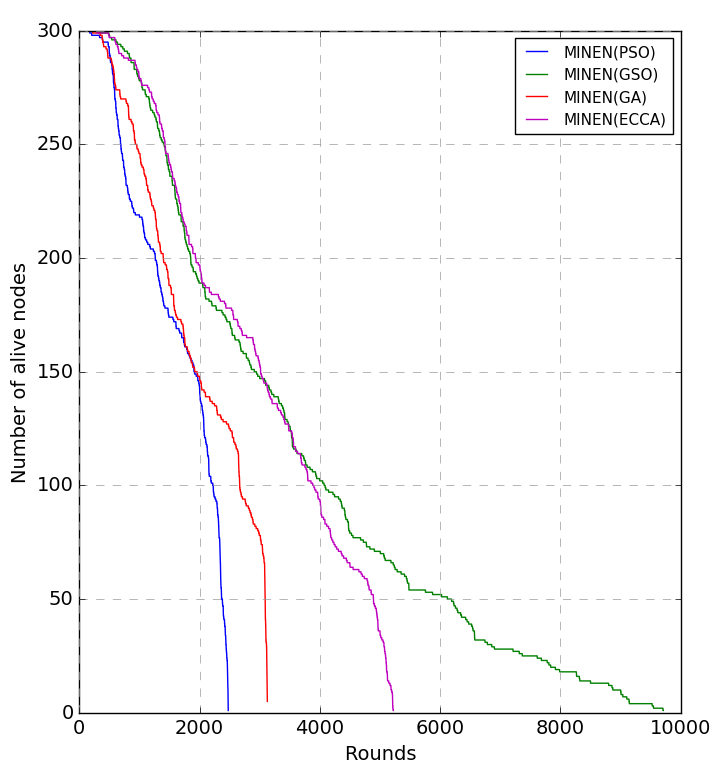}
\caption{}
\end{figure}

Figure 9 depicts the \emph{network coverage} provided by MINEN without sleep scheduling. As can be seen from the figure the maximum simulation area is prominently covered in the shade of red. This signifies that majority of devices in the network are active for 2400 to 3200 rounds of communication. These devices are also more uniformly distributed across the network's geography as compared to LEACH and FCM. Hence MINEN provides superior \emph{network coverage} compared to both LEACH and FCM. Furthermore Figure 10 depicts the \emph{network coverage} of MINEN with GSO sleep scheduling. It should be noted from this figure that the performance of the network is enhanced several folds by the use of sleep scheduling. The prominent blue shades of the coverage graph correspond to 1500-4500 rounds until complete depletion of devices. Hence the network remains active and operational for even more longer periods of time than just using MINEN alone.

In order to evaluate the effectiveness of using GSO with MINEN we perform a comparative analysis of various other sleep scheduling algorithms combined with MINEN. The sleep scheduling algorithms analyzed are : PSO, GA and EECA. We evaluate these algorithms on the basis of \emph{number of active nodes} and \emph{energy dynamics} of the network. Figure 11 plots the energy of the network with number of rounds of communication. As can be observed from the figure, MINEN with GSO enables the network to remain operational for approximately 10,000 rounds of communication. Whereas MINEN with EECA, GA and PSO keep the devices of the network active for about 6000, 3000 and 2500 rounds respectively. The slopes of the energy versus rounds (time) graphs also favor GSO over the other algorithms. PSO has the fastest rate of energy depletion whereas GSO has the least rate of energy depletion. Figure 12 showcases the number of alive nodes in the network versus rounds of communication. It reiterates the inferences made from Figure 11 and hence further consolidates that MINEN when used with GSO performs tremendously better than other state-of-the-art sleep scheduling algorithms.

\section{Conclusion and Future Work}

In this paper, a Minimum Energy (MINEN) routing protocol for IoT-WSNs is proposed. MINEN is a clustering based algorithm which evenly distributes the expense of energy consumption amongst all the devices of the network. This is done using clustering, cluster head rotation and minimization of the energy of sending and receiving messages across links as well as aiding devices with low residual energies. This is achieved by first constructing a DAG where the nodes of the graph are the cluster heads. Then appropriate costs/weights are assigned to the edges keeping in mind the energy required for transmitting/receiving messages over a particular link as well as a factor for a device pair forming a communication link, called Energy spent so far ($E_{sf}$). Inclusion of the $E_{sf}$ factor enforces energy based load balancing across several links in the application. Then, Djikstra's algorithm is employed to find the minimum (energy) cost path of transmitting messages from the sender cluster head to the base station. Furthermore, Genetic Swarm Optimization (GSO) sleep scheduling technique is combined with MINEN to enhance the energy conservation effort. MINEN is seen to perform better when supplemented with GSO as compared to other sleep scheduling techniques. Moreover, MINEN alone outperforms two existing widely used energy efficient routing protocols - LEACH and FCM, in terms of network coverage, number of alive nodes and energy dynamics. 

For future work, the protocol can be expanded to networks where the IoT nodes are all mobile and end-to-end source to destination paths do not exist. Moreover, work can be done to further improve sleep scheduling techniques employed in the paper to achieve even better performance. Work can also be undertaken to analyze other clustering approaches that might give better clusters resulting in more energy conservation.

\ifCLASSOPTIONcaptionsoff
  \newpage
\fi

\end{document}